\documentclass[useAMS,usenatbib]{mn2e}

\usepackage{graphicx}

\title[The structure of Alfv\'en wave driven winds]{On the 
magnetic structure and wind parameter profiles of Alfv\'en wave 
driven winds
in late-type supergiant stars}
\author[D. Falceta-Gon\c{c}alves, A. A. Vidotto \& V. Jatenco-Pereira]{D. Falceta-Gon\c{c}alves$^{1}$\thanks{E-mail: diego@astro.iag.usp.br}, A. A. Vidotto$^{1}$ and V. Jatenco-Pereira$%
^{1}$ \\ $^{1}$Instituto de Astronomia, Geof\'\i sica e Ci\^encias Atmosf\'ericas, \\
Universidade de S\~ao Paulo, Rua do Mat\~ao 1226, CEP 05508-900, S\~ao
Paulo, Brazil}

\begin{document}

\date{ }

\pagerange{\pageref{firstpage}--\pageref{lastpage}} \pubyear{2005}

\maketitle

\label{firstpage}

\begin{abstract}
Cool stars at giant and supergiant evolutionary phases present low velocity
and high density winds, responsible for the observed high mass-loss rates.
Although presenting high luminosities, radiation pressure on dust particles
is not sufficient to explain the wind acceleration process. Among the
possible solutions to this still unsolved problem, Alfv\'en waves are,
probably, the most interesting for their high efficiency in transfering 
energy and momentum to the wind. Typically, models
of Alfv\'en wave driven winds result in high velocity winds if they are not
highly damped. In this work we determine self-consistently the
magnetic field geometry and solve the momentum, energy and mass conservation
equations, to demonstrate that even a low damped Alfv\'en wave flux is able
to reproduce the low velocity wind. We show that the magnetic fluxtubes
expand with a super-radial factor S $>$ 30 near the stellar surface, larger
than that used in previous semi-empirical models. The rapid expansion results
in a strong spatial dilution of the wave flux. We obtained the wind
parameter profiles for a typical supergiant star of 16 M$_\odot$. The wind
is accelerated in a narrow region, coincident with the region of high
divergence of the magnetic field lines, up to 100 km s$^{-1}$. For the
temperature, we obtained a slight decrease near the surface for low damped
waves, because the wave heating mechanism is less effective than the
radiative losses. The peak temperature occurs at $r \simeq 1.5$ r$_0$
reaching 6000 K. Propagating outwards, the wind cools down mainly due to
adiabatic expansion.
\end{abstract}

\pagerange{\pageref{firstpage}--\pageref{lastpage}} \pubyear{2005}

\label{firstpage}


\section{Introduction}

Cool giant and supergiant stars are known to present continuous mass loss
process occuring at high rates, typically $10^{-10}-10^{-5}$ M$_{\odot }$ yr$%
^{-1}$, but in low velocity winds ($u_{\infty }<300$ km s$^{-1}$) (Dupree
1986, Lamers \& Cassinelli 1999). After decades of theoretical and
observational studies, the mechanisms in which the wind acceleration occurs
are still poorly understood. Due to the stellar high luminosity and low
effective temperature several authors have proposed radiativelly dust driven
models to explain the observed wind properties (Liberatore, Lafon \&
Berruyer 2001, Elitzur \& Ivezic 2001, Woitke \& Niccolini 2005). At the
pulsating phase, acoustic waves generate high density shells that could
allow dust to form near the stellar surface. In this scenario, radiative
pressure on grains transfers momentum to these particles being responsible
for their acceleration and, if gas and dust are dynamically well coupled,
grains drag the gas outwards resulting in the mass ejection. However, at
stationary envelopes ($e.g.$ pre-AGB phase) dust driven theoretical models
have lately failed in reproducing the wind properties, mainly because the
dust-gas coupling is not effective (Sandin \& Hofner 2003). Observationally,
Guandalini $et$ $al.$ (2005) found no strong correlation between the mass
loss rates and the luminosities of AGB stars. Their main conclusion is that,
if radiative pressure is important in powering these stellar winds, it must
occurs in addition to other mechanism. Another drawback to the radiation
pressure models is the need for the dust formation region to be close to the
star. Recent high resolution Doppler measurements show that winds are mainly
accelerated near the stellar surface ($r<1.3$ R$_{\ast }$) (Airapetian,
Carpenter \& Ofman 2003), while grains are expected to grow and survive at
even larger distances.

In this sense, another mechanism must be used to accelerate the gas near
surface. The most promising mechanism for the winds of cool stars is the
transfer of momentum and energy to the wind from MHD waves. Alfv\'en
waves driven wind models are known to result in high velocity winds, unlike 
what is measured for giant cool stars, because these MHD waves
are, in general, weakly damped. Hartmann \& MacGregor (1980) showed that it
would be possible to reproduce the observed low wind velocities and the high
mass loss rates of the cool giant and supergiant stars if some kind of wave
damping mechanism is effective at the wind basis ($r < 2$ R$_*$). In that
work, they confirmed this assumption using the ion-friction damping, which
has a damping length proportional to $P^2$, where $P$ is the wave period.
Holzer, Fla \& Leer (1983) questioned this result in terms of an unnatural
fine-tuning for the wave flux period, and argued that stars would, more
reasonably, present a wide variety of wave frequencies depending on the
generation mechanisms, and for most of them the damping would be
uneffective. However, Jatenco-Pereira \& Opher (1989) studied the effects of
both different damping mechanisms and the magnetic field divergence, showing
that the fine-tuning of the wave period is unnecessary since there are several
other damping mechanisms that would act on the frequency spectrum. Also,
they showed that the magnetic field divergent geometry can rapidly dilute
the wave flux and also slow down the wind. Their magnetic field geometry was
based on empirical relations found from observations of the solar wind. This
because in general, in lack of direct measurements of the magnetic field
fluctuations and structure in other stars, we have to simply extrapolate our
knowledge from solar observations.

Parker (1958) proposed that the solar wind is accelerated by the strong thermal
pressure gradient in the transition region, where the gas temperature
increases from $\sim 10^{4}$ K up to $10^{6}$ K. However, it became clear in
the following years that neither the solar radiation, nor acoustic waves
generated in the photosphere, could account for the heating of the coronal
base. The origin of the energy responsible for the heating and acceleration
of the plasma is believed to be from both magnetic field reconnections above
the photosphere (Axford \& McKenzie 1996) and convective motions under
the stellar surface. The convective energy is transferred up to the
atmosphere from the perturbations generated in the field line footpoints
(Cranmer \& van Ballegooijen 2005). These perturbations propagate outwards
as Alfv\'en waves, heating the gas as they are damped (Suzuki \& Inutsuka
2005). Regarding the magnetic field lines, Holzer \& Leer (1980) and
Jatenco-Pereira \& Opher (1989) realized that the super-radial geometry of
the magnetic field funnels at the coronal holes could have a significant
impact on the mass flux and wind speed. Esser $et$ $al.$ (2005), showed that
the models considering highly diverging magnetic funnels explain better the
observed data for the Sun. Also, Tu $et$ $al.$ (2005) established that the
solar wind acceleration initiates in the magnetic funnels at heights lower
than $2.10^9$ cm, coincident with the high divergence region.

In this work, we model the acceleration and heating of a late-type
supergiant stellar wind considering an outward flux of Alfv\'en waves. We
solve the MHD equations to, self-consistently, determine the magnetic field
geometry and the wind temperature, density and velocity profiles. In section
2, we describe the model basic equations. In
section 3 we present the results and the discussions, followed, by 
the work conclusions.

\section{The Model}

The wind equations are based on mass, momentum, energy and magnetic
flux conservation. The first is given by: 
\begin{equation}  \label{mass}
\rho u A(r) = \rho_0 u_0 A(r_0) ,
\end{equation}
where $u$ is the flow velocity, $\rho$ is the gas density and $A(r)$ is the
flow cross-section area at a distance $r$ from the center of the star. The
index ``$0$" indicates the variable is being evaluated at the stellar
surface ($r = r_0$).

Assuming a steady flow, the momentum equation can be written as: 
\begin{eqnarray}  \label{momentum}
\rho (\vec{u}\cdot \vec{\nabla })\vec{u}%
=-\rho \frac{GM_{\ast }}{r^{3}}\vec{r}-\vec{\nabla }P-%
\vec{\nabla }\left( \frac{\langle \left( \delta B\right)
^{2}\rangle }{8\pi }\right)+  \nonumber \\
\frac{1}{4\pi }\left( \vec{B}\cdot \vec{\nabla }%
\right) \vec{B}-\vec{\nabla }\left( \frac{B^{2}}{8\pi }%
\right) \, ,
\end{eqnarray}
where $P = \rho k_B T / m$ is the thermal pressure, $k_B$ is the Boltzmann
constant, $T$ is the gas temperature, $m$ is the mean mass per particle, $G$
the gravitational constant and $\delta B$ the wave magnetic field amplitude.
In Equation (2), the right hand side contains the gravitational force
density and the thermal and wave pressure gradients, respectively. The last
two terms represent the Lorentz force. The wave amplitude ($\delta B$) is
related to the wave energy density ($\epsilon$) by $\epsilon = \langle
(\delta B)^2 \rangle/ (4 \pi)$.

\subsection{Thin fluxtube approximation}

Typically, considering magnetic field strengths $> 1$ G, the wind basis is
characterized by the relation $B^{2}/8\pi \gg P > \rho u^{2}/2$, $i.e.$ the
plasma is magnetically dominated. In this case, if we assume the wind to be
initiated at funnels anchored at the stellar surface, which are surrounded
by a plasma with lower magnetic field strength, the magnetic pressure inside
will push the gas and the funnel field lines will expand. The funnel
cross-section radius ($\mathcal{R}$) will grow super-radially up to a limit
value ($\mathcal{R}_m$). This limiting cross-section radius depends both on
the relation between external and internal magnetic field strengths and on
the filling factor ($\alpha$). As the area increases, the internal magnetic
strength diminishes until the equilibrium between internal and external
magnetic pressures is reached. If the internal magnetic field strength is
much larger than the external, the flux tubes cross-section will depend on
the filling factor only. The filling factor is the ratio between the area of
the stellar surface covered by funnels and the total area. The averaged
maximum area that a funnel could reach would be $A_m = A(r_0)/\alpha$ or, in
terms of the cross section radius: 
\begin{equation}
\mathcal{R}_m = \frac{\mathcal{R}_0}{\alpha^{1/2}} \, .
\end{equation}
For the quiet Sun, the funnels that merge to form the coronal holes cover
about 10\% of the total surface.

To evaluate the tube expansion at the wind basis, we assume the plasma to be magnetically
dominated and the left hand side of Equation (2) may be neglected if
compared to the other terms. Then, by using $\vec{\nabla} \cdot 
\vec{B} = 0$ and a power series expansion method proposed by
Pneuman, Solanki \& Stenflo (1986), we can determine self-consistently the
magnetic field geometry without assuming any empirical function for the
funnel cross-section with distance.

Following Pneuman, Solanki \& Stenflo (1986), using the thin fluxtube
aproximation, Equation (2) and $\vec{\nabla} \cdot 
\vec{B} = 0$ are described near stellar surface by: 
\begin{equation}
4\pi \frac{\partial P}{\partial y} \simeq B_{r}\left( \frac{\partial B_{y}}{%
\partial r}-\frac{\partial B_{r}}{\partial y}\right),
\end{equation}
\begin{equation}
4\pi \left( \frac{\partial P}{\partial r}+\frac{P}{H}\right) \simeq -B_{y
}\left( \frac{\partial B_{y }}{\partial r}-\frac{\partial B_{r}}{\partial y }%
\right),
\end{equation}
and 
\begin{equation}
\frac{1}{y }\frac{\partial }{\partial y }\left( y B_{y }\right) +\frac{%
\partial B_{r}}{\partial r}=0,
\end{equation}
where $H=k_B Tr^{2}/GmM_{\ast }$ is the scale height.

Expanding all variables as power series in $y$ ($i.e.$ along the tube
radius), and neglecting terms of orders higher than 2, Equations (4) $-$
(6) give rise to a differential equation for the fluxtube cross section: 
\begin{eqnarray}
\frac{A(r_{0})}{2H_{0}^{2}}\left[ \frac{\partial ^{2}}{\partial r^{2}}\left( 
\frac{A(r_{0})}{A(r)}\right) -\frac{1}{2A(r)}\frac{\partial }{\partial r}%
\left( \frac{A(r_{0})}{A(r)} \right)\right] =  \nonumber \\
\left( \frac{A(r_{0})}{A(r)} \right) ^{2} \left[ 1-\left( \frac{B_{ext}}{%
B_0}\frac{\left( 1-\alpha \right) }{\left( \frac{A(r_{0})}{A(r)}-\alpha
\right) }\right) ^{2}\right] +2\beta \frac{P(r)}{P(r_{0})},
\end{eqnarray}
where $\beta = 4 \pi P(r_0)/B^2(r_0)$, $\alpha$ is the filling factor and $%
B_{ext}$ is the magnetic field strength outside the fluxtube. In the
following calculations we fixed its value to be $10^{-3}B(r)$.

To simplify the set of equations, we will define the funnel area expansion
as a function of radial distance by: 
\begin{equation}
A\left( r\right) =A\left( r_{0}\right) \left( \frac{r}{r_{0}}\right) ^{S} \,
,
\end{equation}
where $S$ is the expansion index, which is super-radial (S $>$ 2) at the
wind basis up to the merging radius when $S$ becomes 2. $S$ is determined
from Equation (7).

\subsection{The energy equation}

Falceta-Gon\c calves \& Jatenco-Pereira (2002) showed that, even though
presenting low effective temperatures, the temperature gradients at the wind
basis of late-type supergiant stars could play an important role in accelerating
the gas, as occurs in the Sun. In their model, the temperature profile was
assumed to be a $r$ dependent function, obtained from observational data.
The wind temperatures in giant cool stars are expected to increase from the
photosferic value up to $10^4$ K, much lower than what is observed for the
Sun, but still important in the wind acceleration if the gradient occurs in
small lengthscales. From their calculations, Falceta-Gon\c calves \&
Jatenco-Pereira (2002) also showed that the thermal pressure is even more important
than radiation pressure at the non-pulsating phase of these objects.

In a consistent model, to avoid assuming any empirical function for the 
magnetic field geometry and to
determine the wind temperature at each wind position ($r$), we have to solve
the energy equation, which is determined from the balance between wave
heating and the adiabatic expansion and radiative coolings (Hartmann,
Edwards \& Avrett 1982; Vidotto \& Jatenco-Pereira 2006). Thus, neglecting conduction, we write the energy
equation as: 
\begin{equation}  \label{energy}
\rho u \frac{d}{dr} \left( \frac{u^2}{2} + \frac52 \frac{k_B T}{m} - \frac{G
M_\star}{r} \right) + \frac{u}{2} \frac{d \epsilon}{dr} = (Q - P_R) \, ,
\end{equation}
where $\epsilon = \left< \left( \delta B \right)^2 \right> / \left( 
4\pi \right)$ is the wave 
energy density, which is described below (Sec. 2.3), 
$Q$ is the wave heating rate, $i.e.$ the rate at which the gas is
being heated due to dissipation of wave energy, and $P_R$ is the radiative
cooling rate, both in erg~cm$^{-3}$~s$^{-1}$. The wave heating can be
written as: 
\begin{equation}
Q = \frac{\epsilon}{L} (u + v_A)
\end{equation}
and the radiative cooling is given by: 
\begin{equation}
P_R = \Lambda \, n_e \, n_H \, ,
\end{equation}
where $n_e$ is the electron density, $n_H$ is the hydrogen density and $%
\Lambda$ is the radiative loss function. Here, we adopt the $\Lambda$
function given by Schmutzler \& Tscharnuter (1993) and calculate $n_e$ with
the modified Saha equation given by Hartmann \& MacGregor (1980).

\subsection{Wave energy density}

The wave energy density at each step may be calculated, using a WKB
aproximation, from the wave action conservation. Under this assumption, the
wave energy density is dissipated as follows: 
\begin{equation}
\epsilon = \epsilon_0 \frac{M_0}{M} \left( \frac{1+M_0}{1+M} \right)^2 \exp %
\left[ - \int_{r_0}^{r} \frac{1}{L} \, dr^{\prime}\right] \, ,
\end{equation}
where $M=u/v_A$ is the Alfv{\'e}n-Mach number, $v_A = (B / \sqrt{4 \pi \rho})
$ the Alfv{\'e}n speed and $L$ the wave damping length. Also, the wave flux (%
$\phi_A$) at $r_0$ is evaluated by (Jatenco-Pereira \& Opher 1989): 
\begin{equation}
\phi_{A_{0}} = \epsilon_0 v_{A0} \left( 1 + \frac32 M_{0} \right) \, .
\end{equation}

Actually, the WKB aproximation is not valid when the perturbation wavelength
is much larger than the gradients lengthscales of the system, $e.g.$ when $%
\lambda >$ $L$. As a consequence, part of the wave flux is reflected and the
Alfv\'en waves propagating on opposite directions decay as the generated beat
waves interact with the gas particles. Davila (1985) pointed out that, in
the Sun, the gradients of the coronal parameters are high enough to make the
WKB aproximation not applicable. However, Usmanov $et$ $al.$ (2000) found good
agreement between the observational data and models with this approach. In
this model, we used Equation (12) assuming then linear perturbations, however
included non-linear effects in the damping length function ($L_{NL}$). The
non-linear damping length is given by (Jatenco-Pereira \& Opher 1989): 
\begin{equation}
L_{NL}\ =L_{0}\left( \frac{v_{A}}{v_{A0}}\right) ^{4}\frac{\langle
(\delta v)^{2}\rangle _{0}}{\langle (\delta v)^{2}\rangle }%
\left( 1+M\right) \, ,
\end{equation}
where $\langle(\delta v)^{2}\rangle$ is the averaged squared
perturbation velocity amplitude and $L_0$ is the
damping length at the wind basis, which is mainly dependent on the assumed
wave frequency spectrum (Lagage \& Cesarsky 1983). Here, we will let it as a
free parameter.

\subsection{The wind equations}

Finally, once the magnetic field structure is determined, we evaluate 
the temperature and velocity of the wind along the magnetic field. 
Using the Equation (8) and considering the magnetic
flux conservation, Equations (1), (2) and (9) in the 
radial direction are reduced to (Vidotto \& Jatenco-Pereira 2006): 

\begin{equation}  \label{energy2}
\frac{dT}{dr} = \frac23 \frac{T}{r} \left[ \frac{r(Q-P_R)}{\rho u (k_B T /m)}
- \left( S + \frac{r}{u} \frac{du}{dr} \right) \right] \, ,
\end{equation}
and 
\begin{eqnarray}  \label{momentum2}
\frac{1}{u} \frac{du}{dr} \left[ u^2 - \frac53 \frac{k_B T}{m} - \frac{%
\langle (\delta v)^2 \rangle}{4} \left( \frac{1+3M}{1+M} \right) \right] = 
\nonumber \\
= \frac{S}{r} \left[ \frac53 \frac{k_B T}{m} - \frac23 \frac{r(Q-P_R)}{S
\rho u} - \frac{G M_\star}{rS} + \right.  \nonumber \\
\left. + \frac{\langle (\delta v)^2 \rangle}{2LS} r + \frac{\langle (\delta
v)^2 \rangle}{4} \left( \frac{1+3M}{1+M} \right) \right]\, .
\end{eqnarray}

Equations (13) $-$ (16) fully describe the wind parameters and the magnetic
field geometry under the given assumptions. In the next section, we show the
main results by applying these equations in a typical cool supergiant star
and compare them with previous works.

\section{Results and Discussions}

For the last decades, the validity of a wind model was constrained in
reproducing only both the terminal velocity and the mass-loss rate of a
given star. As a consequence, a number of accelerating mechanisms were found
in accordance to observations of these constrains. However, with the high
resolution observations and more sensitive intruments the parameters radial
profiles will become measurable and will be decisive on the modelling
choice. Here we determine the velocity, density and temperature profiles
and discuss their dependence on initial assumptions.

We applied the described model on a cool supergiant star with $M_*$ = 16 M$%
_\odot$, $r_0 = 400$ R$_\odot$, $\rho_0 = 10^{-13}$ g cm$^{-3}$, $B_0 = 10$
G and $T_0 = 3500$ K. We also assumed a filling factor $\alpha = 0.1$,
according to solar observations.

\begin{figure}
{\includegraphics[scale=.35]{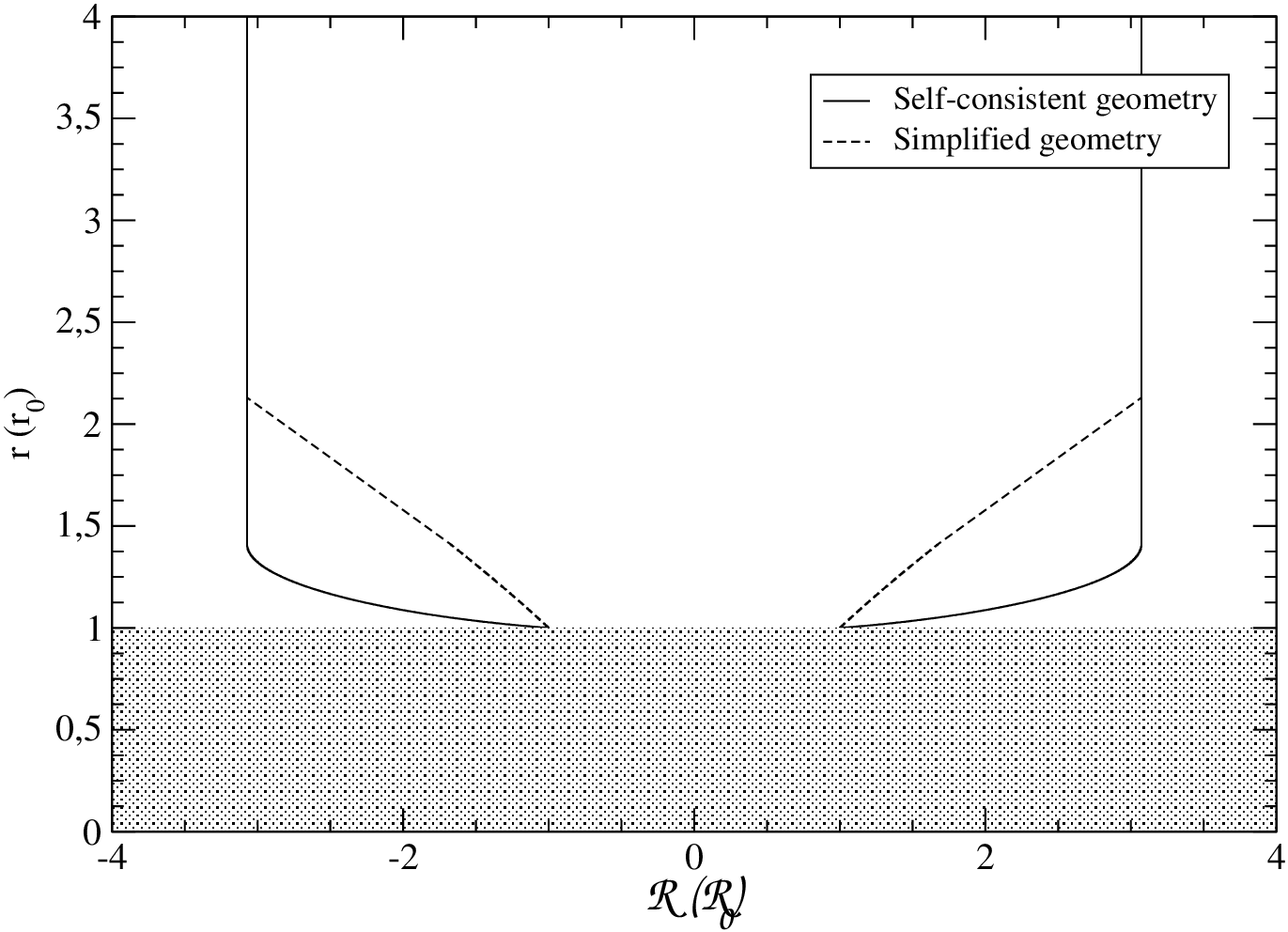}}
\caption{The magnetic field structure for a constant super-radial index 
$S = 5$ (dashed line) and that determined self-consistently in our model (solid 
line).}
\label{fig1}
\end{figure}

\subsection{Wind profiles}

To study the importance of the damping length on the wind profiles we
performed calculations setting the initial wave flux as $\phi_{A0} = 3.10^{7}
$ erg cm$^{-2}$ s$^{-1}$, and varied $L_0$ as 0.2, 1.0 and 5.0 r$_0$. The
magnetic field structure is shown in Figure (1). The solid line represents
the solution of Equation (7) for the given stellar parameters, compared to
a simple constant expansion factor $S = 5$ (dashed line), as used by
Falceta-Gon\c calves \& Jatenco-Pereira (2002). Comparing both curves we
find that the self-consistent geometry is more divergent near stellar
surface, and the flux tubes are expected to merge at lower distances, in
agreement with more refined models for the Sun (Esser $et.$ $al.$ 2005). 
We obtained an initial value of $S(r_0) \simeq 35$, that decreases with 
distance to the stellar surface. From the assumed $\alpha$, we found that 
the neighbour fluxtubes merge at $r = 1.41$ r$_0$. At this point, with a 
sudden decrease of $S$ from its super-radial value to $S = 2$ (radial), 
the wind properties change and it is noticeable as breaks in the temperature, 
density and velocity profiles.

We obtained the same 
geometry for the following calculations considering different 
initial parameters. This because the super-radial magnetic field 
geometry occurs mainly at the wind
basis, where the magnetic pressure is highly dominant. In this sense, 
we expect the expansion index ($S$) to be independent on the other 
initial parameters such as the wind velocity, the wave flux 
densities and the damping length. 

In Figure (2) we compare the wind velocity profiles for the different
initial damping lengths. As expected, the larger the damping length the
higher is the wind velocity. However, for previous works that do not take
into account any high initial divergence, the wind velocity is even larger.
The obtained divergence result in a fast dilution of the wave energy density
near the stellar surface. As a consequence, the lack of wave flux at larger
distances results in lower velocities. This result shows that even low damped
wave fluxes can be responsible for low terminal velocities, depending on the
field divergence.

\begin{figure}
{\includegraphics[scale=.35]{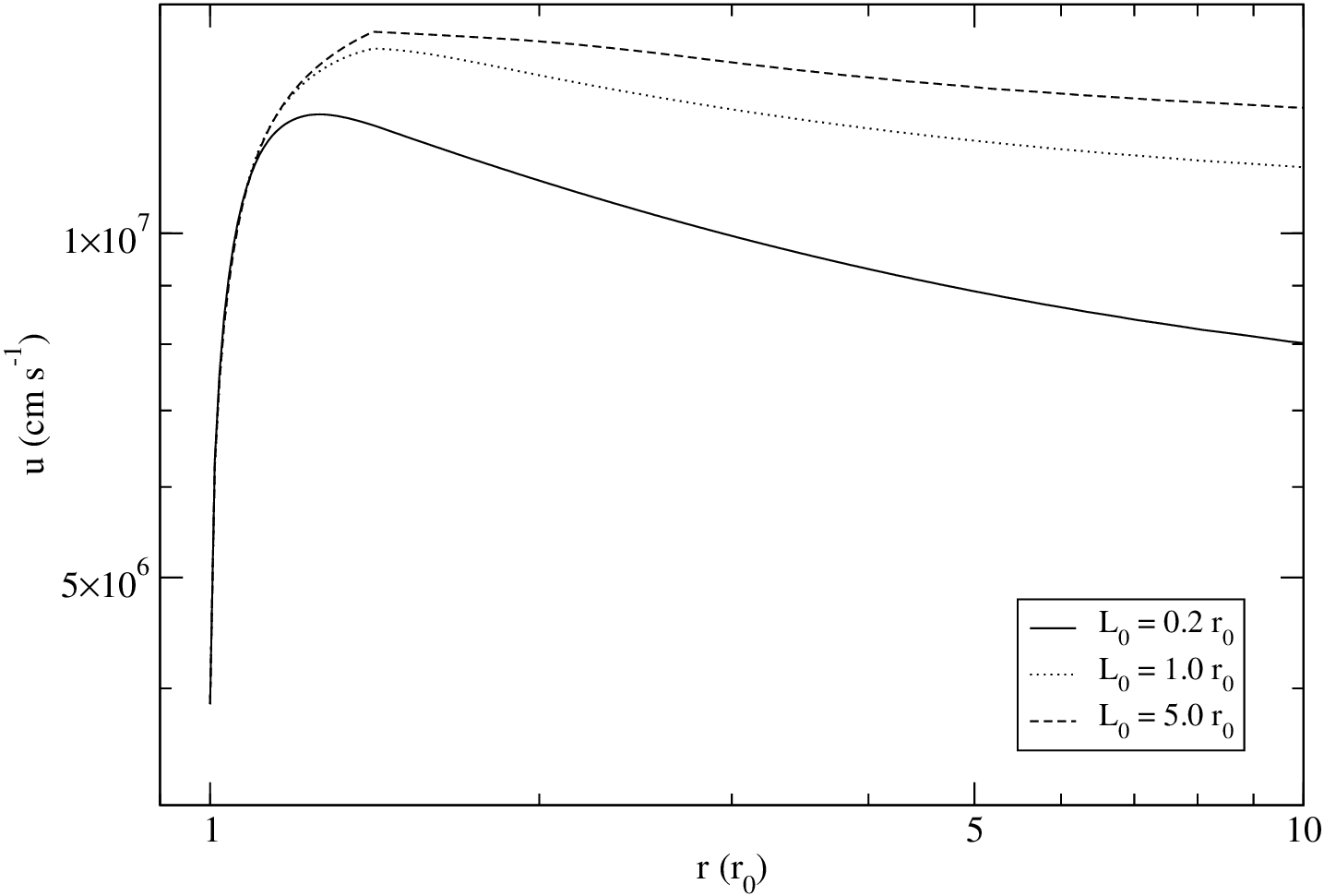}}
\caption{Wind velocity profiles obtained for the initial damping 
lengths $L=0.2$ r$_0$ (solid line), $L=1.0$ r$_0$ (dotted line) and 
$L=5.0$ r$_0$ (dashed line) as function of the distance. }
\label{fig2}
\end{figure}

In Figures (3) $-$ (5) we present the gas mass density, wave energy density
and the temperature, respectively, for the same calculations. For the gas
density, all three lines show the same behaviour at $r <$ 1.5 r$_0$ indicating
that the flux tube divergence is the dominant factor if comparing to the gas
expansion due to acceleration. This is an interesting result since, typically,
the density profile is believed to be closely related to the velocity
profile only. In Figure (4), we notice that, in despite of what was obtained
in previous works, the wave energy density is weakly dependent on the
initial damping length at the wind basis. This mainly because, again, the
field divergence is dominant on the dilution process. On the other hand, for
distances larger than the flux tubes merging position the wave damping is
dominant. In Figure (5), the temperature profiles show
different features. For the given stellar parameters, if $L_0=5.0$~r$_0$, the
wave heating is less effective and the radiative losses added to the
expansion cooling result in a negative temperature gradient, which occurs in
a very narrow region until the density becomes low enough for the radiative
losses become less important. For highly damped waves, the wave heating
dominates the radiative losses and the temperature gradient is positive near
the surface. Another interesting feature is the maximum temperature
position. Since the temperature gradient is mainly dependent on the wave
damping ($via$ $Q$ parameter), we notice that the lower is the damping
length, the closer to the stellar surface the maximum temperature occurs.

\begin{figure}
{\includegraphics[scale=.35]{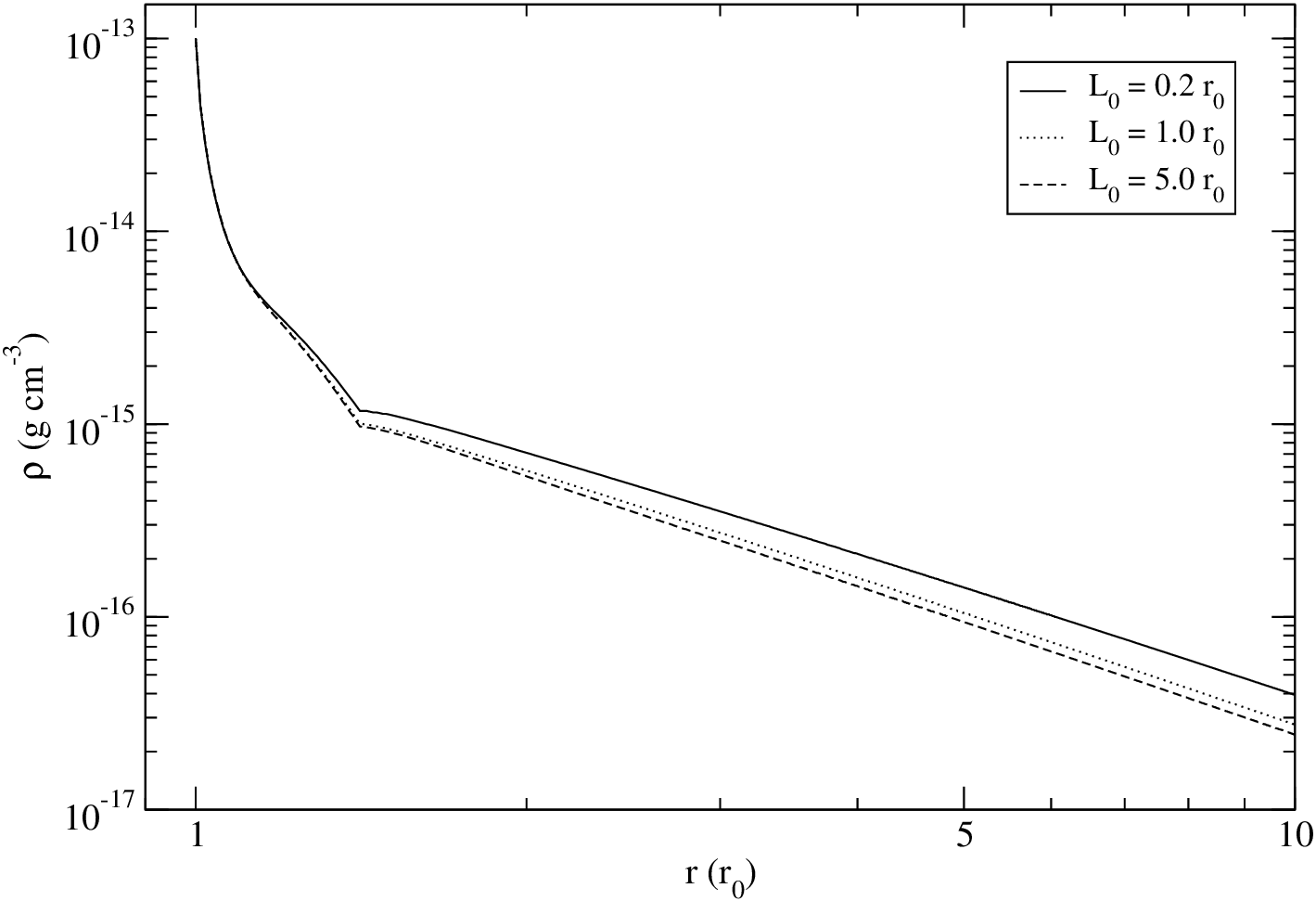}}
\caption{Wind density profiles obtained for the initial damping 
lengths $L=0.2$ r$_0$ (solid line), $L=1.0$ r$_0$ (dotted line) and 
$L=5.0$ r$_0$ (dashed line) as function of the distance.}
\label{fig3}
\end{figure}

\begin{figure}
{\includegraphics[scale=.35]{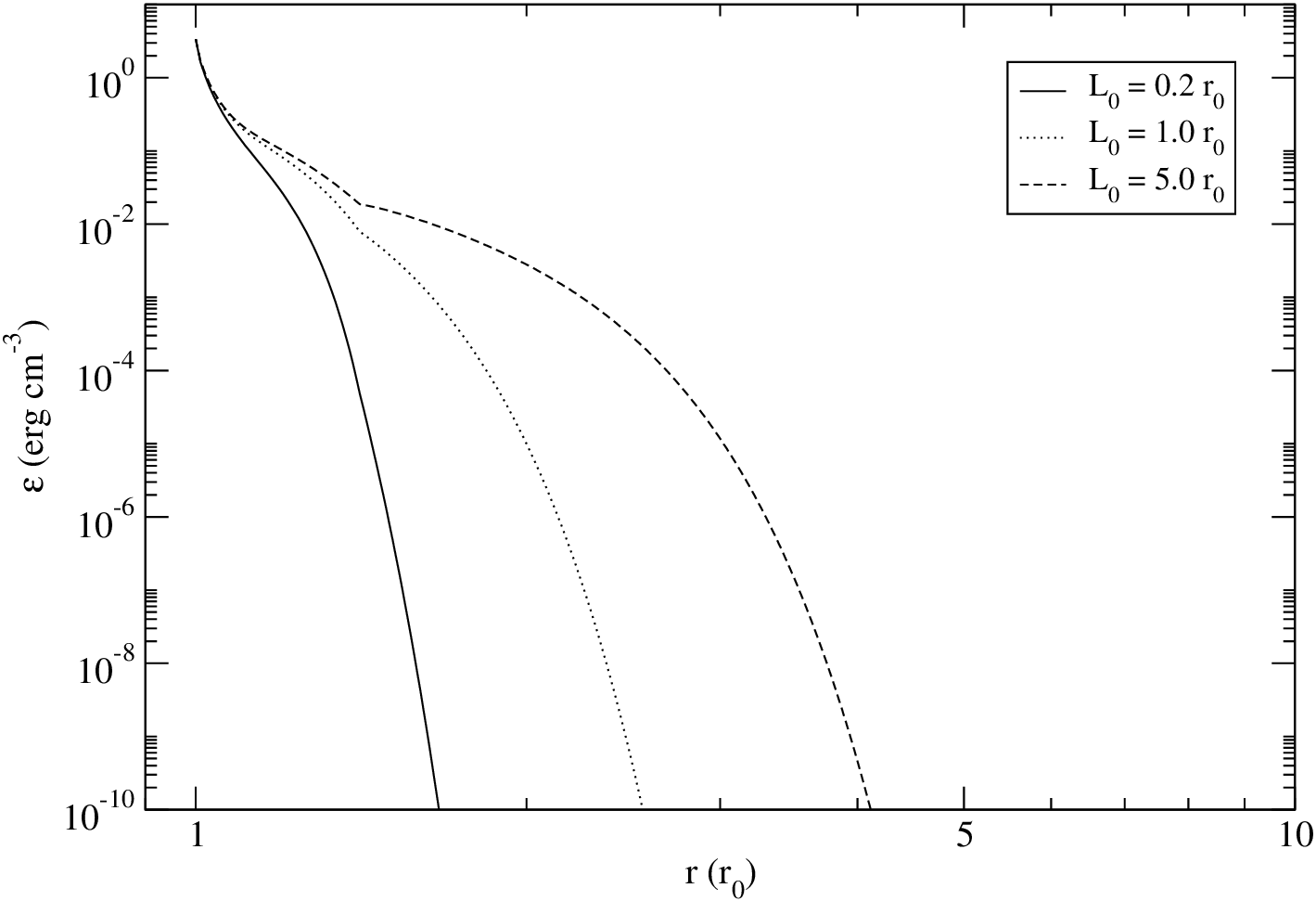}}
\caption{Wave energy density profiles obtained for the initial damping 
lengths $L=0.2$ r$_0$ (solid line), $L=1.0$ r$_0$ (dotted line) and 
$L=5.0$ r$_0$ (dashed line) as function of the distance.}
\label{fig4}
\end{figure}

\begin{figure}
{\includegraphics[scale=.35]{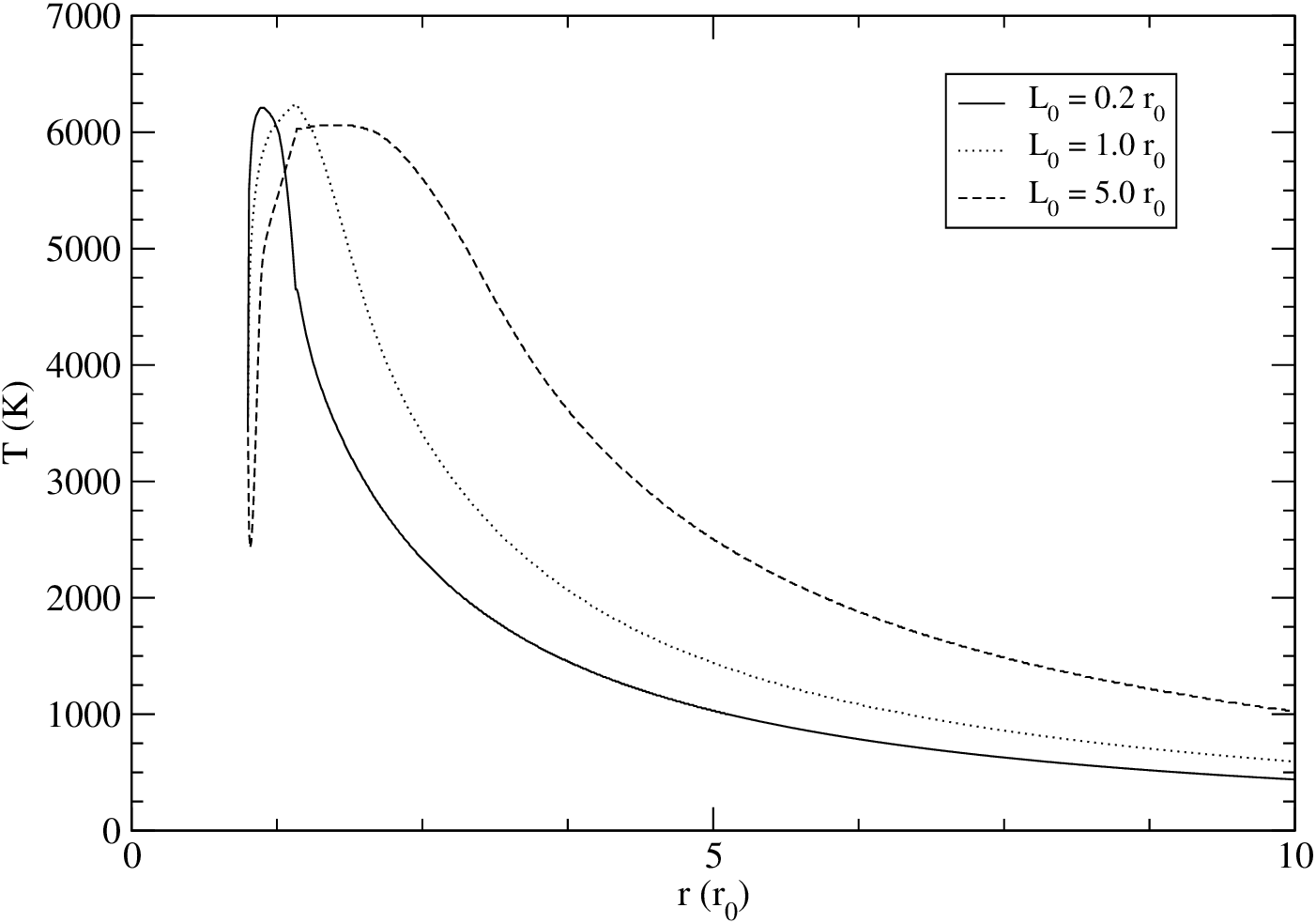}}
\caption{Wind temperature profiles obtained for the initial damping 
lengths $L=0.2$ r$_0$ (solid line), $L=1.0$ r$_0$ (dotted line) and 
$L=5.0$ r$_0$ (dashed line) as function of the distance.}
\label{fig5}
\end{figure}

Recently, Zeeman splitting of H$_2$O masers observations led Vlemmings,
Diamong \& van Langevelde (2002) to infer magnetic field strengths on
supergiant stellar surfaces of $B_0 > 100$ G. Also, theoretical dynamo
models for AGB stars reveal that the planetary nebulae shapes can be
explained by magnetic field strengths of $\sim 200$ G (Blackman $et.$ $al.$
2001). For this reason, we also perfomed numerical calculations for $B_0
= 10$ and 100 G, setting the initial parameters for a low damped wave flux ($%
L_0 = 5$ r$_0$). We also assumed the same wave amplitude relative to the
magnetic field for both cases.

The velocity profile for both initial magnetic field strengths are shown in
Figure (6). For the larger magnetic field strength ($B_0 = 100$ G) we
obtained the higher terminal velocity ($u > 200$ km s$^{-1}$). This effect
is mainly due to the higher wave energy flux assumed at the wind basis to 
accomplish the same relative wave amplitude. 

The temperature profiles, shown in Figure (7), also present features
dependent on the magnetic field strength. As for the velocity, a high wave
energy flux results in a high temperature, at least near the stellar surface
due to the higher energy density. 

\begin{figure}
{\includegraphics[scale=.35]{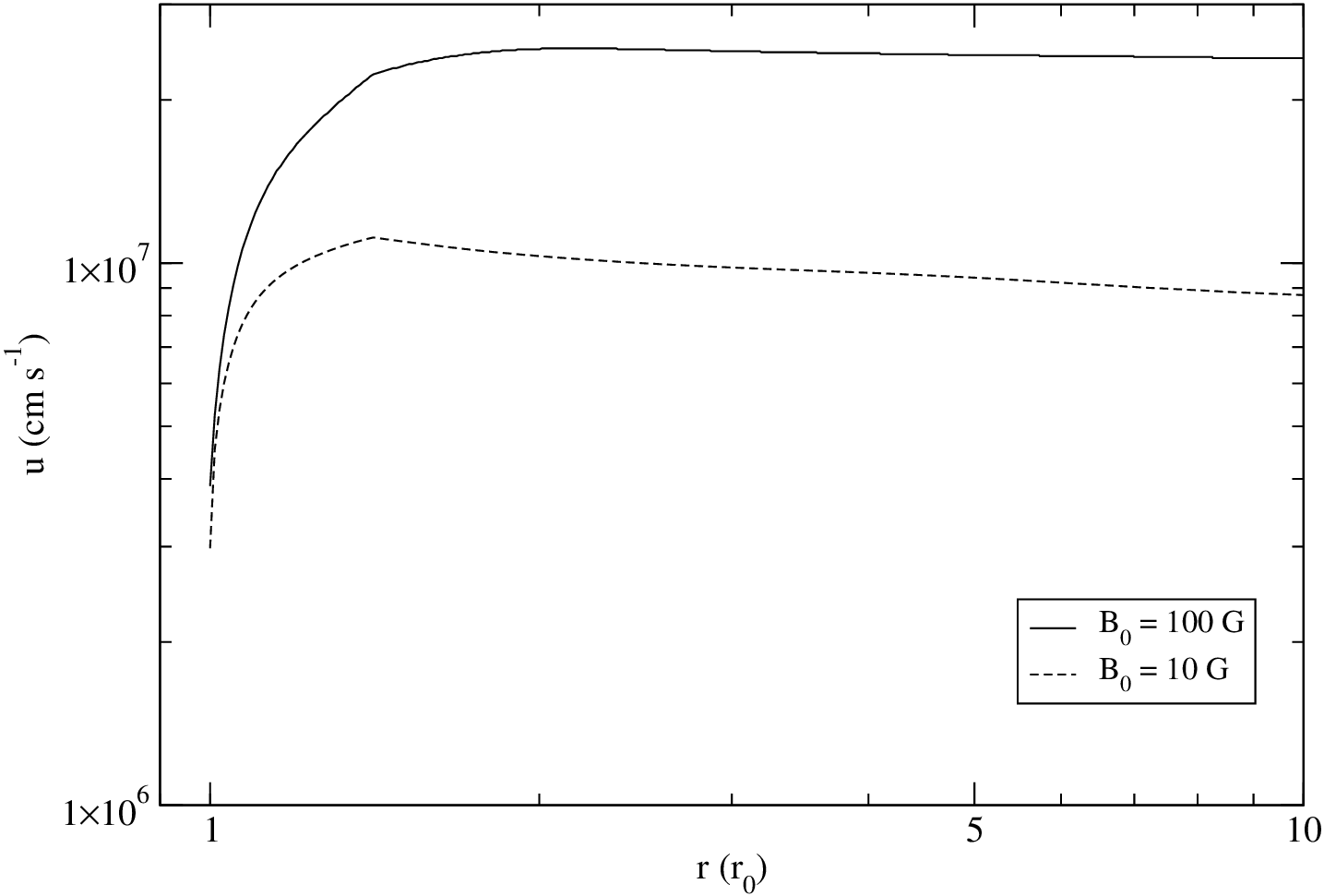}}
\caption{Wind velocity profiles obtained for initial magnetic field 
strengths $B_0 = 10$ G (dotted line) and $B_0 = 100$ G 
(solid line) as function of the distance.}
\label{fig6}
\end{figure}

\begin{figure}
{\includegraphics[scale=.35]{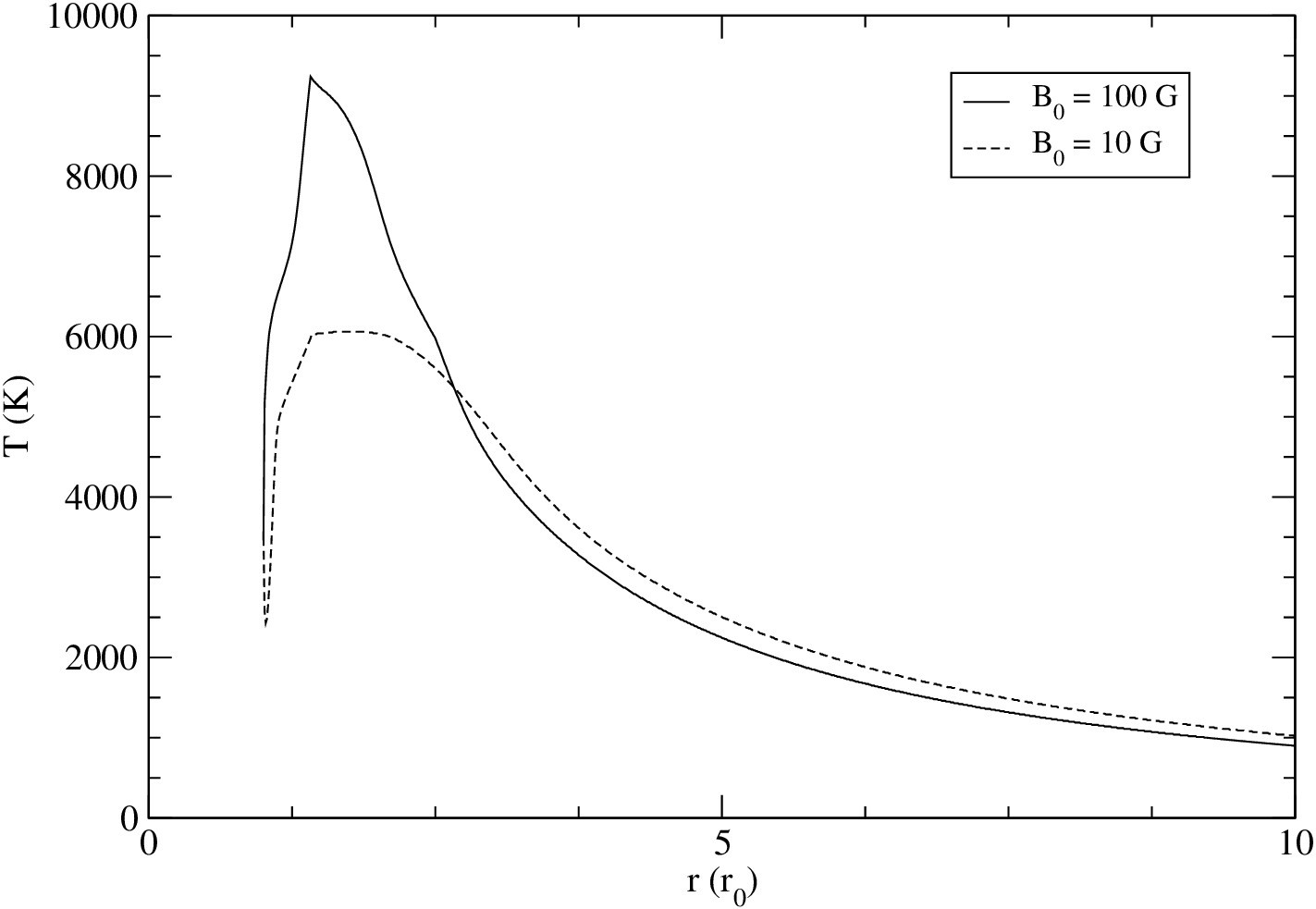}}
\caption{Wind temperature profiles obtained for initial magnetic field 
strengths $B_0 = 10$ G (dotted line) and $B_0 = 100$ G 
(solid line) as function of the distance.}
\label{fig7}
\end{figure}

\subsection{Best-fitting model}

For stars with parameters similar to those used in the previous subsection,
observational data reveal typical mass-loss rates of $\dot{M} \simeq 10^{-7} 
- 10^{-6}$
M$_\odot$ yr$^{-1}$ and terminal velocities of $u \simeq 70$ km s$^{-1}$.
Unfortunately, the available data are limited in spatial resolution for most
of the stars, and it is not possible to fit the complete radial profiles.

Assuming a surface magnetic field strength $B_0 = 10$ G, and a low damped
wave flux ($L_0 = 5$ r$_0$), it was possible to reproduce both the wind
terminal velocity and the mass loss rate using an Alfv\'en waves flux of $%
\phi_{A0} = 10^7$ erg cm$^{-2}$ s$^{-1}$ at the wind basis. This value
corresponds to a wave amplitude of $\langle \left( \delta
B\right)^{2}\rangle^{\frac{1}{2}} \simeq$ 3.10$^{-2}$ $B_0$, which is very
plausible for a turbulent medium as that at the stellar surface.

The velocity and temperature profiles for this case are shown in Figures (8)
and (9), respectively. The velocity profile reveals a peak of $u > 100$ km s$%
^{-1}$ at $r < 2.0$ r$_0$, and slightly decreases for larger distances until
reaching the observed value. The temperature profile presents an initial
negative gradient reaching temperatures $T < 2500$ K in a narrow region.
Also, near $r = 1.5$ r$_0$ the temperature reaches the maximum value of $%
\sim 6000$ K. For $r > 3.0$ r$_0$, where the wave heating and the radiative
losses are low, the temperature decreases due to the adiabatic expansion.

\begin{figure}
{\includegraphics[scale=.35]{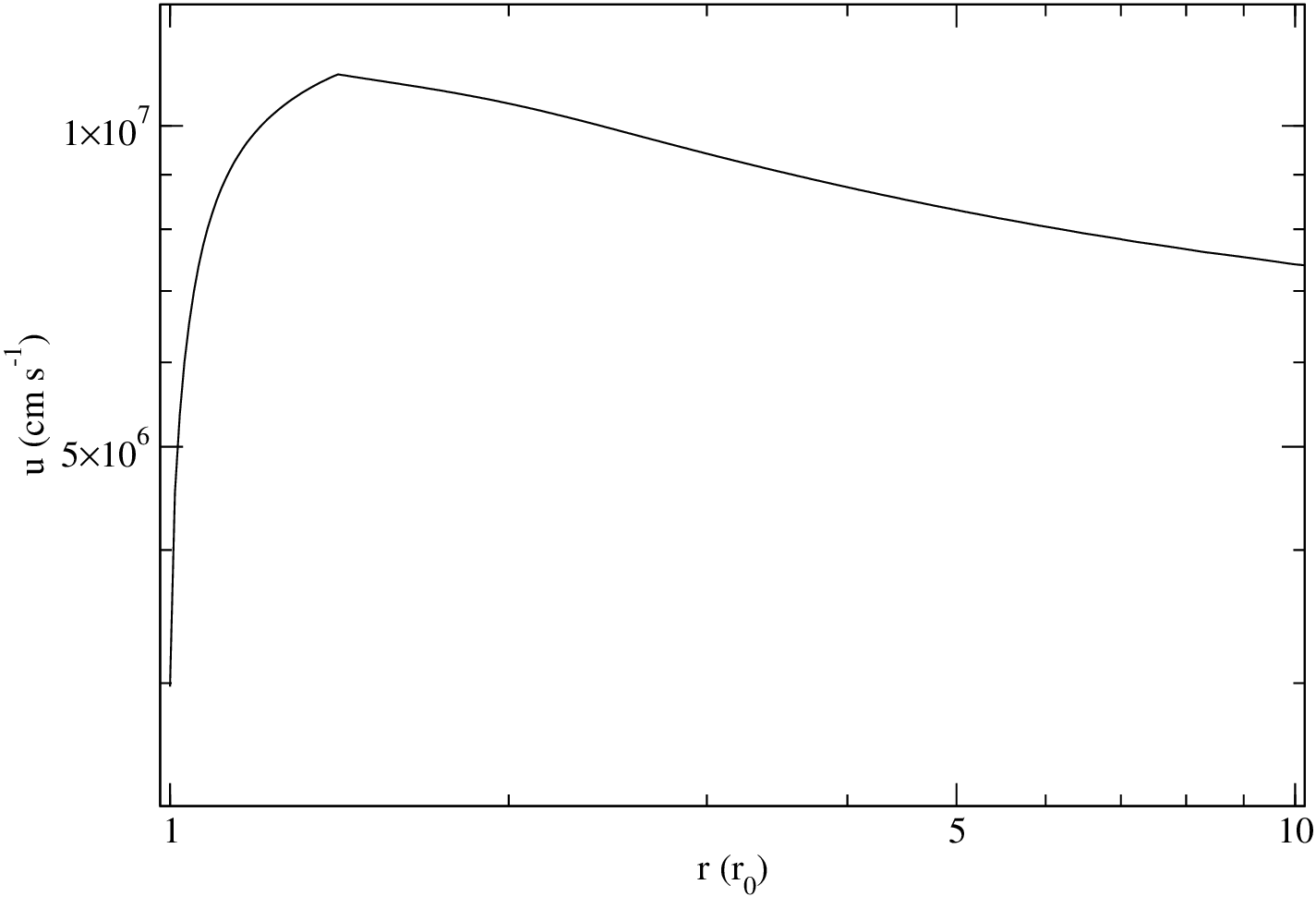}}
\caption{The wind velocity profile for the best fitting parameters in the 
case of low damped waves. Here we assumed $B_0 = 10$ G, $L_0 = 5.0$ r$_0$ and 
$\phi_{A0} = 10^7$ erg cm$^{-2}$ s$^{-1}$.}
\label{fig8}
\end{figure}

\begin{figure}
{\includegraphics[scale=.35]{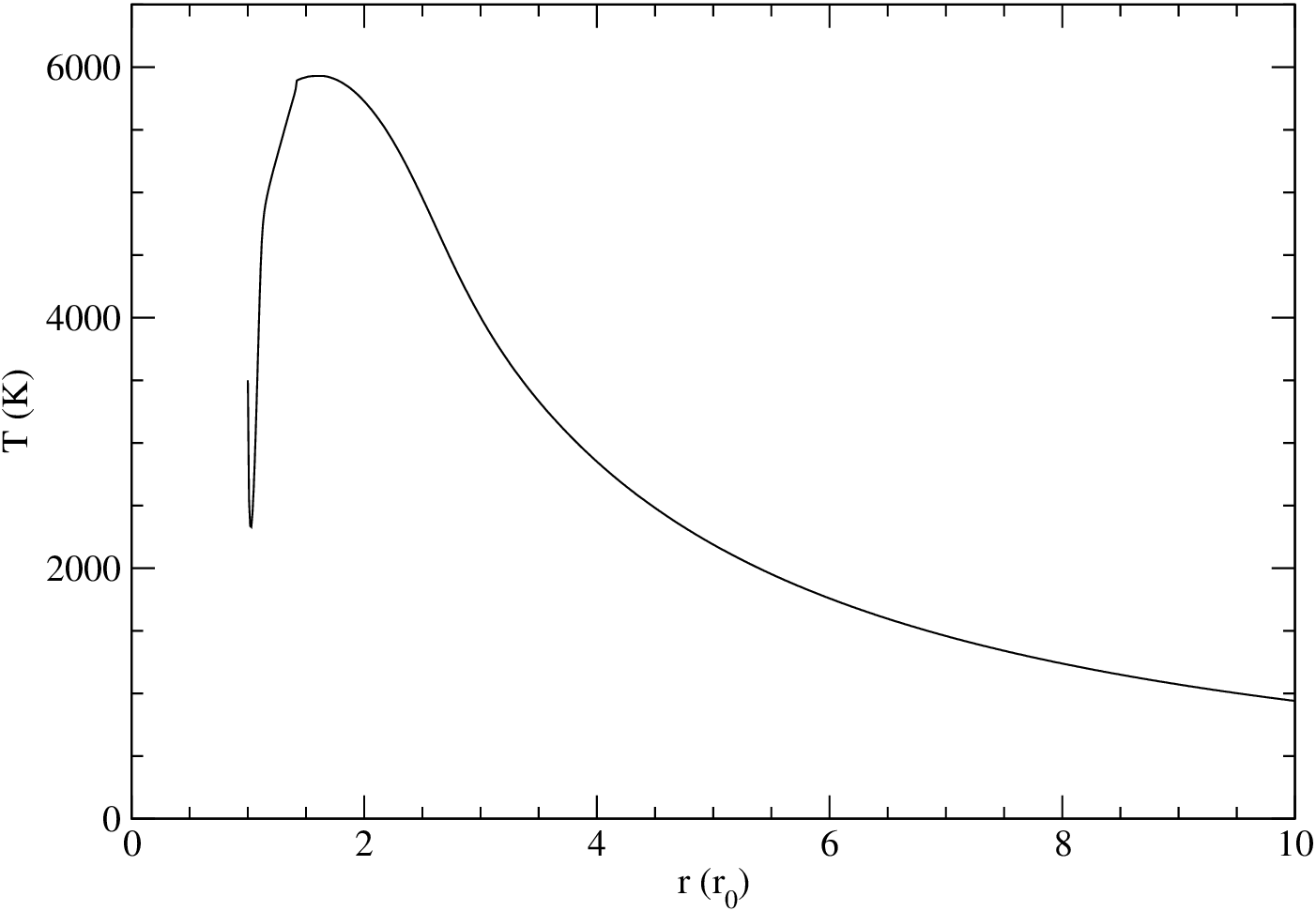}}
\caption{The wind temperature profile for the best fitting parameters in the 
case of low damped waves. Here we assumed $B_0 = 10$ G, $L_0 = 5.0$ r$_0$ and 
$\phi_{A0} = 10^7$ erg cm$^{-2}$ s$^{-1}$.}
\label{fig9}
\end{figure}

\section{Conclusions}

We propose a self-consistent wind model to determine the 
parameters profiles for a supergiant late-type star. 
To determine the magnetic field geometry we used 
an expansion method over the wind physical parameters as proposed by Pneuman, 
Solanki \& Stenflo (1986). Near surface, the magnetic 
pressure inside the flux tubes are higher 
than that of the surrounding medium, forcing the field lines to curve. 
We found an initial super-radial expansion factor $S > 30$ at the wind basis, much 
higher than the value used by previous authors that included empirical 
relations to account for the magnetic field geometry based on solar observations.

Considering a supergiant late-type star with $M_*$ = 16 M$
_\odot$, $r_0 = 400$ R$_\odot$, $\rho_0 = 10^{-13}$ g cm$^{-3}$, $B_0 = 10$
G and $T_0 = 3500$ K, we obtained the wind velocity, density and temperature 
profiles. Typically, Alfv\'en wave driven winds result in high velocity 
winds ($u > 100$ km s$^{-1}$), unless some strong wave damping mechanism 
takes place at the wind basis. We showed that this conclusion is 
correct in the case of low divergent magnetic field structures. In this work, the 
strong divergence is responsible for a rapid wave spatial dilution near 
surface, resulting in a lower wind velocity even for low damped waves. 

We reproduced both the typical mass-loss rate 
($\dot{M} \simeq 10^{-7} - 10^{-6}$ M$_\odot$ yr$^{-1}$) and 
terminal velocity ($u \simeq 70$ km s$^{-1}$) observed for these objects, 
assuming a weakly damped ($L_0 = 5.0$ r$_0$) Alfv\'en wave flux of $\phi_{A0} = 
10^7$ erg cm$^{-2}$ s$^{-1}$. The velocity profile reveals an efficient 
acceleration at $r < 1.5$ r$_0$, reaching the maximum value $\sim 100$ km 
s$^{-1}$. In this region the wind is mainly accelerated by the wave energy 
density and the thermal pressure gradients. Afterwards, the absence of the wave 
acceleration and the cooling gas result in a decrease of the velocity to the 
observed values. For the temperature, assuming a weakly 
damped wave flux, the 
radiative losses and the expansion cooling are dominant near surface, 
and the temperature gradient is initially negative. The temperature falls 
to $\simeq 2500$ K in a sharp region and then, as density decreases as the 
wind accelerates and the flux tube expands, it increases up to $\simeq 6000$ K 
at $r < 2.0$ r$_0$. For higher distances, where the radiative losses 
are low and the wave heating is no longer effective, the temperature 
decreases mainly due to the adiabatic expansion.

\section*{Acknowledgments}

D. Falceta-Gon\c calves and A. A. Vidotto thank the Brazilian agency
FAPESP for the finantial support (04/12053-2 and 04/13846-6). V. 
Jatenco-Pereira thanks CNPq for the finantial support (304523/90-9).

\label{lastpage}

\end{document}